 \documentstyle[12pt,twoside,fleqn,espcrc1,epsfig]{article}

\newcommand{\AmS}{{\protect\the\textfont2
  A\kern-.1667em\lower.5ex\hbox{M}\kern-.125emS}}
\newcommand{\beq}{\begin{eqnarray}}
\newcommand{\eeq}{\end{eqnarray}}

\title{Single Transverse Spin Asymmetry in $p^\uparrow p \to \pi X$ and
$ep^\uparrow \to \pi X$} 

\author{Yuji Koike\address[MCSD]{Department of Physics, Niigata University, 
        Niigata 950-2181, Japan} 
}       

\begin{document}

\maketitle

\begin{abstract}
Single transverse spin asymmetry for the pion production 
in $pp$ and $ep$ collisions is 
studied in the framework of collinear factorization.
The soft-gluon pole contributions with twist-3 distribution and 
fragmentation functions are identified and its characteristic features are
discussed.
\end{abstract}

\vspace{0.5cm}

In this report we discuss the single transverse-spin asymmetry $A_N$ 
for the pion production with large transverse momentum 
in $pp^\uparrow\to\pi(\ell) X$ and $p^\uparrow e \to \pi(\ell) X$
relevant for RHIC-SPIN, HERMES and COMPASS experiments {\it etc}. 
According to the QCD factorization theorem,
the polarized cross section for $pp^\uparrow \to \pi X$  
consists of three twist-3 contributions:
\beq
&(A)&\qquad G_a(x_1,x_2)\otimes q_b(x')\otimes D_{c}(z)\otimes
\hat{\sigma}_{ab\to c},\nonumber\\
&(B)&\qquad  \delta q_a(x)\otimes E_b(x_1',x_2') \otimes D_{c}(z)\otimes 
\hat{\sigma}_{ab\to c}',\nonumber\\
&(C)&\qquad \delta q_a(x)\otimes q_b(x') \otimes \widehat{E}_{c}(z_1,z_2)
\otimes\hat{\sigma}_{ab\to c}'',\nonumber
\eeq
where 
the functions $G_a(x_1,x_2)$, $E_b(x_1',x_2')$ 
and $\widehat{E}_{c}(z_1,z_2)$
are the twist-3 quantities representing, respectively, the
transversely polarized distribution, the unpolarized distribution, and
the fragmentation function for the pion. 
Other functions are twist-2: $q_b(x')$ is the unpolarized quark or gluon
distribution,
$\delta q_a(x)$ is the transversity distribution in $p^\uparrow$, and
$D_c(z)$ is the fragmentation function for the pion.
$a$, $b$ and $c$ stand for the parton's species, sum over which
is implied.  $\delta{q}_a$, $E_b$ and $\widehat{E}_c$ are chiral-odd.
$\hat{\sigma}_{ab\to c}$ {\it etc} are the partonic hard cross section
producing large transverse momentum of the pion.    
Corresponding to the above
(A) and (C), the polarized cross section for $ep^\uparrow \to \pi X$
(final electron is not detected) receives two twist-3 contributions:
\beq
&(A')&\qquad G_a(x_1,x_2)\otimes 
D_{a}(z)\otimes\hat{\sigma}_{ea\to a}, \nonumber\\
&(C')&\qquad \delta q_a(x)\otimes\widehat{E}_{a}(z_1,z_2)\otimes
\hat{\sigma}_{ea\to a}'.\nonumber
\eeq
The (A) and (B) contributions for $pp^\uparrow\to \pi X$
have been analyzed in \cite{QS99} and \cite{KK00}, respectively, and
it has been shown that (A) gives rise to large $A_N$ at large $x_F$ 
as observed in E704, and (B) is negligible in all kinematic region. 
Here we extend the study to the (C) term
at RHIC energy and also the asymmetry in $ep$ collision.

The transversely polarized twist-3 distribution 
$G_{F}^a(x_1,x_2)$ relevant to $G_a$ in the (A) and (A') terms
is given in \cite{KK00}, and 
the twist-3 fragmentation function 
$\widehat{E}_{F}^c(z_1,z_2)$ for $\widehat{E}_c$
in (C) and (C') is defined in \cite{Koike02}.
According to a detailed analysis, $G_F^a$ and $\widehat{E}_F^c$ 
contribute to the
cross section at $x_1=x_2=x$ and $z_1=z_2=z$, respectively (soft-gluon-pole
contribution).  In particular, the derivarives $(d/dx)G_F^a(x,x)$ and
$(d/dz)\widehat{E}_F^c(z,z)$ appear in the cross section formula. 
In the large $x_F$ region, i.e. for the production of pion 
in the forward direction of the polarized nucleon, 
the main contribution comes from the large-$x$ and large-$z$ region
of distribution and fragmentation functions, respectively. 
Since $G_F^a$ and $\widehat{E}_F^c$ behaves as $G_F^a(x,x)\sim (1-x)^\beta$
and $\widehat{E}_F^c(z,z)\sim (1-z)^{\beta'}$ with $\beta$, $\beta'>0$,
$|(d/dx)G_F^a(x,x)|\gg |G_F^a(x,x)|$,
$|(d/dz)\widehat{E}_F^c(z,z)| \gg |\widehat{E}_F^c(z,z)|$ at
large $x$ and $z$.  In particular, the valence component of
$G_F^a$ and $\widehat{E}_F^c$ dominate in this region.
We thus keep only the valence quark contribution for the derivative
of these soft-gluon pole functions (``valence-quark soft-gluon
approximation'') for the $pp$ collision.  For the $ep$ case,
we include all the soft-gluon pole contribution.

In general
$A_N$ is a function of 
$S=(P+P')^2\simeq 2P\cdot P'$,
$T=(P-\ell)^2\simeq -2P\cdot \ell$ and
$U=(P'-\ell)^2\simeq -2P'\cdot \ell$ where $P$, $P'$ and $\ell$
are the momenta of $p^\uparrow$, unpolarized $p$ (or $e$), 
and the pion respectively. 
In the following we use
$S$, $x_F = {2\ell_{\parallel}\over \sqrt{S}} = {T-U\over S}$ and
$x_T = {2\ell_{T}\over \sqrt{S}}$ as independent variables.
The cross section formula for the (C) term 
is given in \cite{Koike02}.

\begin{figure}[htb]
\setlength{\unitlength}{1cm}
\begin{picture}(14,4.7)(0,-0.25)
\psfig{file=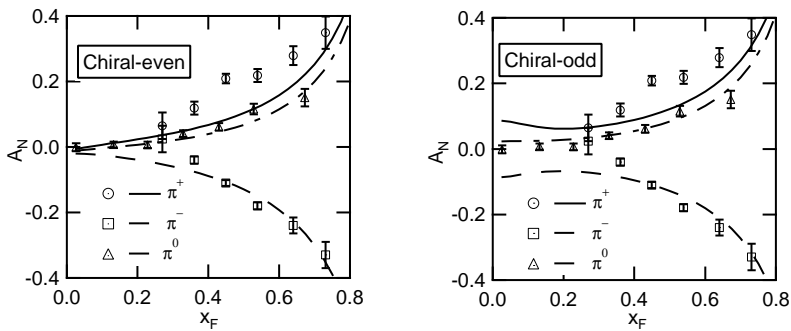,height=5cm}
\end{picture}\par
\vspace*{-1.2cm}
\caption{$A_N^{pp}$ at $\sqrt{S}=20$ GeV and $\ell_T=1.5$ GeV
together with E704 data.}
\end{figure}

\vspace{-0.3cm}

To estimate the (C) contribution,
we introduce a simple model ansatz as
$\widehat{E}_F^a(z,z)=K_a D_{a}(z)$ 
with a flavor dependent factor $K_a$.
Assuming that the (C) is the sole origin of $A_N$,
$K_a$'s are determined to be $K_u=-0.11$ and $K_d=-0.19$ 
so that it can approximately reproduce $A_N^{pp}$ observed
in E704 data at $\sqrt{S}=20$ GeV and $\ell_T=1.5$ GeV.  
We refer the readers to \cite{KK00}
for the adopted distribution and fragmentation functions.
The result for $A_N^{pp}$ from the (C) (chiral-odd) term
is shown in 
Fig. 1 in comparison with the (A) (chiral-even) contribution.
The magnitude of the (A) term in Fig. 1 is also determined
with the ansatz $G_F^a(x,x)=K'_a q_a(x)$ ($K_u'=-K_d'=0.07$)
so that it can approximately reproduce E704 data by itself.
Recall that 
our valence-quark soft-gluon approximation is valid at large $x_F$.
Both effects give rise to $A_N^{pp}$ similar to the E704 data.
The origin of the growing
$A_N$ at large $x_F$ is
(i) large partonic cross sections ($\sim 1/\hat{t}^2$ term)
and (ii) the derivative of the soft-gluon pole functions.
  
With the parameters $K_a$ and $K_a'$ fixed,
$A_N^{pp}$ at RHIC energy ($\sqrt{S}=200$ GeV)
is shown in Fig.2 at $l_T=1.5$ GeV.  
Both (A) and (C) contributions give slightly 
smaller $A_N^{pp}$ than the STAR data
reported at SPIN2002.
Fig. 3 shows the $\ell_T$-dependence of $A_N^{pp}$
at $\sqrt{S}=200$ GeV and
$x_F=0.6$, indicating quite
large $l_T$ dependence in both (A) and (C) contributions
at $1< l_T < 4$ GeV region, which is typical to the twist-3 effect
($\sim O(1/\ell_T)$).  Polarized cross section receives
$O(\ell_T/S)$ contribution as well, so that the $A_N$ tends to become 
more stable at larger $\ell_T$.  
 
\begin{figure}[htb]
\setlength{\unitlength}{1cm}
\begin{minipage}[t]{9cm}
\begin{picture}(9,4.7)(0.7,-0.5)
\psfig{file=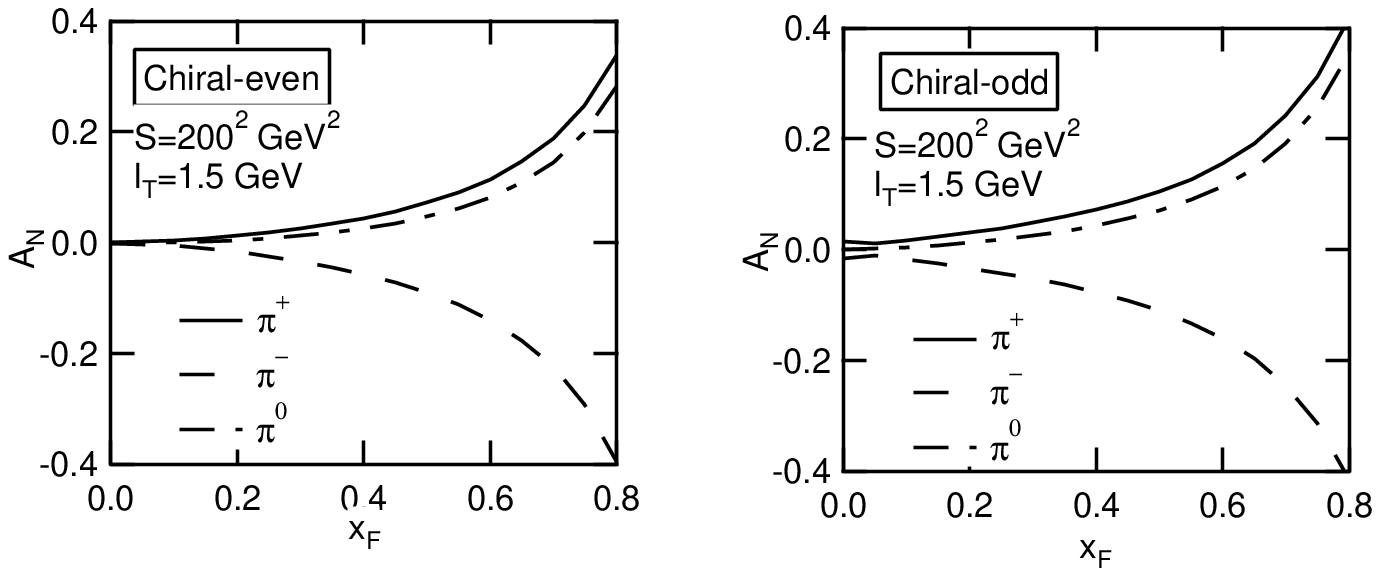,height=5cm}
\end{picture}\par
\vspace*{-1.5cm}
\caption{$A_N^{pp}$ at $\sqrt{S}=200$ GeV and $\ell_T=1.5$ GeV.}
\end{minipage}
\hfill
\begin{minipage}[t]{4.8cm}
\begin{picture}(4.8,4.7)(0.5,-0.5)
\psfig{file=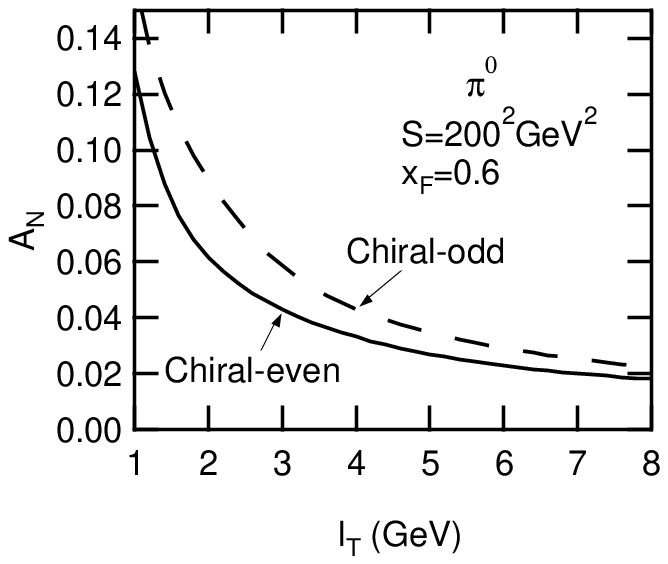,height=5cm}
\end{picture}\par
\vspace*{-1.5cm}
\caption{$\ell_T$ dependence of $A_N^{pp}$ at $\sqrt{S}=200$ GeV and 
$x_F=0.6$.}
\end{minipage}
\end{figure}

\vspace{-0.4cm}

We next discuss the asymmetry $A_N^{ep}$ for $p^\uparrow e\to \pi(\ell)X$
where the final electron is not observed.  
In our $O(\alpha_s^0)$ calculation, 
the exchanged photon remains highly virtual
as far as the observed $\pi$ has
a large transverse momentum $\ell_T$ with respect to the $ep$ axis.
Therefore experimentally one needs to integrates only over those 
virtual photon events ($Q^2 > (S/2)[1-x_F/\sqrt{x_F^2+x_T^2}]$)
to compare with our formula.

\begin{figure}[htb]
\setlength{\unitlength}{1cm}
\begin{picture}(14,4.7)(0,-0.5)
\psfig{file=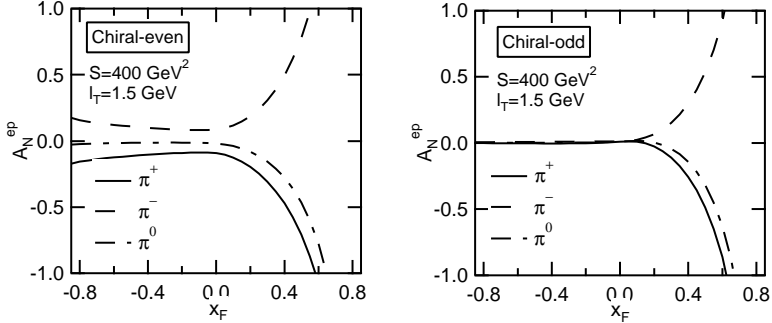,height=5cm}
\end{picture}\par
\vspace*{-1.5cm}
\caption{$A_N^{ep}$ at $\sqrt{S}=20$ GeV and $\ell_T=1.5$ GeV.}
\end{figure}

\vspace{-0.4cm}

Using the twist-3 distribution and fragmentation functions 
used to describe $pp$ data, we show in Fig. 4 
$A_N^{ep}$ corresponding to
(A')(chiral-even) and (C')(chiral-odd) contributions.
Remarkable feature of Fig. 3 is that in both chiral-even and
chiral-odd contributions (i) the sign of $A_N^{ep}$
is opposite to the sign of $A_N^{pp}$ and (ii) the magnitude of $A_N^{ep}$
is much larger than that of $A_N^{pp}$, in particular, at large $x_F$,
and it even overshoots one.
(In our convention, $x_F >0$ 
corresponds to the production of $\pi$ in the forward hemisphere of
the initial polarized proton both in $p^\uparrow p$ and $p^\uparrow e$ case.)
The origin of these features can be traced back to the color factor in
the dominant diagrams for the {\it twist-3 polarized} cross sections
in $ep$ and $pp$ collisions.
Of course, the asymmetry can not exceeds one, and thus our model 
estimate needs to be modified.  
First, the applied kinematic range
of our formula should be reconsidered: 
$A_N$ at $\ell_T$ smaller than a few GeV may not be ascribed to
the twist-3 effect.
Second, our model ansatz of $G_F^a(x,x)\sim q^a(x)$ and 
$\widehat{E}_F^a(z,z)\sim D^a(z)$ is not appropriate at $x\to 1$ and
$z\to 1$, respectively.  The derivative of these functions,
which is important for the growing $A_N^{pp}$ at large $x_F$, 
eventually
leads to divergence of $A_N^{pp}$ at $x_F\to 1$ as $\sim 1/(1-x_F)$.

As a possible remedy to
this pathology we tried the following.
For the (A) and (A') (chiral-even)
contribution, as an example, 
we have a model $G_F^a(x,x)\sim q_a(x)\sim_{x\to 1} (1-x)^{\beta_a}$
where $\beta_u=3.027$ and $\beta_d=3.774$ in the GRV distribution we adopted.
Tentatively we shifted $\beta_{u,d}$ as $\beta_a\to\beta_a(x)=\beta_a+x^3$,
which suppresses the divergence of $A_N$ at $x_F\to 1$ but still
causes rising behavior of $A_N$ at large $x_F$.  
The result for $A_N$ with this modification of $G_F(x,x)$
is shown in Fig. 5 ($A_N^{ep}$), Fig. 6 ($A_N^{pp}$ at E704 energy) and
Fig. 7 ($A_N^{pp}$ at RHIC energy).  
This avoids 
overshooting of one in $A_N^{ep}$ but reduces $A_N^{pp}$, typically 
by factor 2, which makes
the deviation from E704 and
STAR data larger.  A similar modification of $\widehat{E}_F(z,z)$ 
in the (C) and (C') (chiral-odd) contribution also leads to the same change
in $A_N$.  

\begin{figure}[htb]
\setlength{\unitlength}{1cm}
\begin{minipage}[t]{5cm}
\begin{picture}(5,4.7)(0.4,-0.5)
\psfig{file=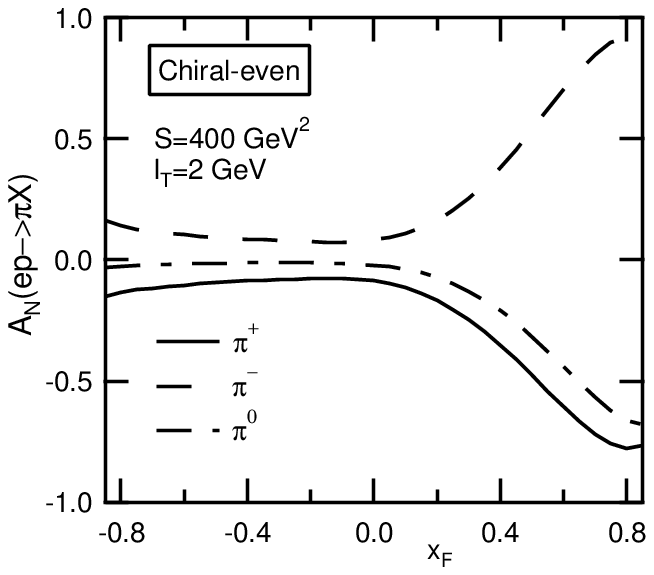,height=5cm}
\end{picture}\par
\vspace*{-1.5cm}
\caption{$A_N^{ep}$ with a modified $G_F$.}
\end{minipage}
\hfill
\begin{minipage}[t]{5cm}
\begin{picture}(5,4.7)(0.4,-0.5)
\psfig{file=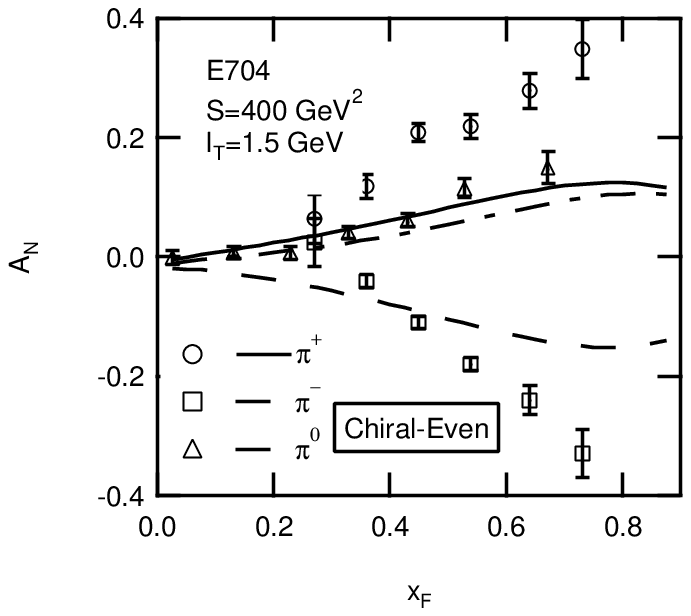,height=5cm}
\end{picture}\par
\vspace*{-1.5cm}
\caption{$A_N^{pp}$ with a modified $G_F$ at $\sqrt{S}=20$ GeV.} 
\end{minipage}
\hfill
\begin{minipage}[t]{5cm}
\begin{picture}(5,4.7)(0.4,-0.5)
\psfig{file=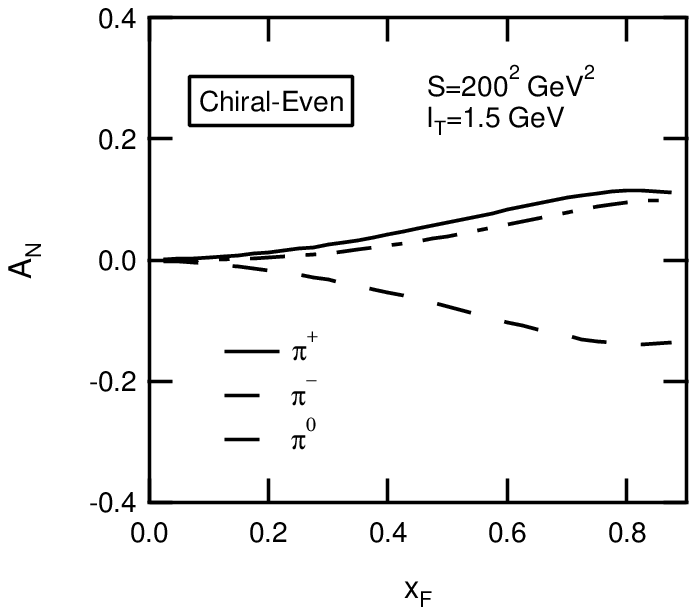,height=5cm}
\end{picture}\par
\vspace*{-1.5cm}
\caption{$A_N^{pp}$ with a modified $G_F$ at $\sqrt{S}=200$ GeV.}
\end{minipage}
\end{figure}

\vspace{-0.4cm}

To summarize we have studied the $A_N$ for pion
production in $pp$ and $ep$ collisions.  
We have pointed out that both (A)(chiral-even) and (C)(chiral-odd) 
((A') and (C') for $ep$
collisions) contributions
can be equally important sources for the asymmetry.  
Although our approach
provides a systematic framework for the large $\ell_T$ production,
applicability of the formula to the currently available low $\ell_T$
data still needs to be tested, in particular,  
detailed comparison of $\ell_T$-dependence and the global analysis
of $pp$ and $ep$ data
is necessary.

\vspace{0.5cm}
\noindent
{\bf Acknowledgement:}  This work is supported in part 
by the Grant-in-Aid for Scientific Research of Monbu-Kagaku-sho.



\begin{thebibliography}{9}
\bibitem{QS99} J. Qiu and G. Sterman, Phys. Rev. {\bf D59} (1999) 014004.
\bibitem{KK00} Y. Kanazawa and Y. Koike, Phys. Lett. {\bf B478} (2000) 121;
{\bf B490} (2000) 99.
\bibitem{Koike01} Y. Koike, hep-ph/0106260 (Proceedings of DIS2001, Bologna, 
Italy, April, 2001.)
\bibitem{Koike02} Y. Koike, hep-ph/0210396 (Proceedings of SPIN2002,
Long Island, USA, Sep. 2002)
\end{thebibliography}
\end{document}